\documentstyle[12pt]{article}
\def\be{\begin{equation}}
\def\ee{\end{equation}}
\begin{document}

\setcounter{footnote}0
\begin{center}
\hfill AIV - 98/II\\
\vspace{0.3in}
\bigskip
\bigskip
\bigskip
{\Large \bf  Near  Anti-de Sitter Geometry

\bigskip

and

\bigskip

Corrections to the Large N Wilson Loop
}
\date{}
\bigskip

\bigskip

\bigskip

{\Large Anastasia Volovich}

\bigskip

\begin{center}
{\it
Department for Theoretical Physics, Moscow State University, 119899;\\
L. D. Landau Institute for Theoretical Physics, Kosygina, 2, 117940;
\\ Moscow, Russia
\\e-mail: nastya@itp.ac.ru}
\end{center}

\bigskip

\end{center}

\begin{abstract}

Within recent Maldacena's proposal to relate gauge theories 
in the large N limit to the supergravity in the $AdS$ background 
and recipe for calculation the Wilson loop, we compute corrections 
to the energy of quark/anti-quark pair in the large N limit.
\end{abstract}
\newpage

Recently, Maldacena has put forward a remarkable
proposal \cite{Mald1} relating the gauge theories in the large N limit to the
supergravity in the anti-de Sitter background times some compact manifold.
This conjecture was further studied in a large number of papers
\cite{SS}-\cite{Kuk} appeared lately. 
A precise recipe for computing CFT observables
in terms of $AdS$ space was suggested in
\cite{GKP, Wit2}. It was proposed that the correlation functions
in CFT are given by the dependence of the supergravity action
on the asymptotic behavior at the boundary.

In \cite{RY},\cite{Mald2} it was suggested that in order to
compute the
vacuum expectation value of the Wilson loop operator
\be
\label{WL}
W({\cal C})= \frac{1}{N} Tr P \exp{i \int_{{\cal C}} A_{\mu} dx^{\mu}},
\ee
one has to consider a string theory partition function
on a certain background with the worldsheet
of the string ending on the contour on the boundary of AdS.
In particular, it was shown that the energy of the quark-antiquark pair
for 4d ${\cal N}=4$ SYM exhibits a Coulomb like behavior.

In this note we will consider  corrections to
the energy of the quark/anti-quark pair.
As we know in QCD the leading term in the 1/N expansion
describes the spectrum of  free mesons and the first 1/N corrections
describe the interaction of mesons \cite{Wit-old}.

\bigskip
\bigskip

The main example of \cite{Mald1} is the consideration 
of 4d ${\cal N}=4$ SYM which is conjectually
dual to IIB superstring in the $AdS_5 \times S^5$ background.
Let us start with the supergravity solution
describing N parallel D3-branes in IIB theory:
\be
ds^2=f^{-1/2}(dt^2 +dx_1^2+dx_2^2 +dx_3^2)+
f^{1/2}(dr^2 + r^2 d\Omega^2),
\label{D3}
\ee
where
\be
f=1+\frac{4\pi g N {\alpha'}^2}{r^4}.
\label{f}
\ee
If one defines the new variable
$U=\frac{r}{\alpha'}$ and takes $\alpha' \to 0$
limit, keeping $U$ fixed, then
neglecting 1 in the harmonic function, ({\ref{D3})
describes $AdS_5 \times S^5$ \cite{Mald1}.
Let us take  into account
the first $1/N$ corrections to this, then   the
near anti-de Sitter metric can be written as follows
\be
ds^2=\alpha'
[\frac{U^2}{R^2}(1-\frac{\alpha'^2 U^4}{2R^4})
(dt^2 +dx_1^2+dx_2^2 +dx_3^2)+
R^2(1+\frac{\alpha'^2 U^4}{2 R^4})(\frac{dU^{2}}{U^2} +  d\Omega_5^2)],
\label{metric}
\ee
where $R=(4 \pi g N)^{1/4}$ corresponds to the radius of
the $AdS_5$ and $g \sim g^2_{YM}$.

The  proposal for computation the vacuum expectation values
of the Wilson loop operators (\ref{WL}) made in \cite{Mald2} is
that
\be
<W({\cal C})> \sim exp(-S)
\ee
where S is the area of a string worldsheet
which at the boundary of $AdS$ describes the loop $\cal{C}$,
from which one has to  subtract the contribution
of the mass of the W-boson, so that it is finite.
It was shown that the energy of quark and anti-quark pair
is inverse proportional to the distance between them.
In the case of finite temperature one has to consider the near-extremal
solution, which was discussed in \cite{Wit3}-\cite{BISY}.

Let us calculate the energy of the quark/anti-quark
pair taking into account the $\alpha'^2$ corrections,
corresponding to the rectangular Wilson loop $L \times T$,
where L is the distance between quark and antiquark and T is large.
We as in \cite{Mald2} have to start with the ordinary Nambu-Goto
action for the string
\be
\label{sa}
S=\frac{1}{4\pi \alpha'}
\int dt d\sigma
\sqrt{det (G_{MN}\partial_\alpha X^M \partial_\beta X^N)},
\ee
where $G_{MN}$ is the background (\ref{metric}). Under the embedding
\be
X^1=t,~~~X^2=x,~~~X^5=U(x),
\ee
the action takes the form:
\be
\label{action}
S=\frac{T}{2 \pi} \int dx \sqrt{(\partial_x U)^2 +\frac{U^4}{R^4}(1-
\alpha '^2 \frac{U^4}{R^4})}.
\ee
Assumption  that the  correction 
\be
\label{rest}
\alpha '^2 \frac{U^4}{R^4}<<1
\ee
is small will lead us to a cutoff, $U<<U_{max}$.
The first integral for such an action is
\be
\frac{\frac{U^4}{R^4}(1-{\alpha'}^2 \frac{U^4}{R^4})}
{ \sqrt{(\partial_x U)^2+
\frac{U^4}{R^4}
-\alpha'^2\frac{U^8}{R^8}}}=
\frac{U_0^2}{R^2} \sqrt{1-{\alpha'}^2 \frac{U_0^4}{R^4}},
\label{1int}
\ee
where $U_0$ is the minimum value of $U$.
Now as in \cite{Mald2} we first have to determine the dependence of $U_0$
from $L$ and then calculate $E(L)$. From (\ref{1int}) we get
\be
\label{L1}
\frac{LU_0}{2R^2}=
 \sqrt{1-\lambda^4}
\int_{1}^{\nu/\lambda}
\frac{dy}
{ y^2\sqrt{(1-\lambda ^4 y^4)(y^4-1)(1-\lambda ^4(y^4+1))}},
\ee
where
\be
\lambda =\sqrt{\alpha '}U_0/R.
\ee
and $\nu$ is a number defined by the cut-off $U_{max}$.
One has here  a natural cutoff  $U_{max}=R/\sqrt{\alpha'}-U_0
\approx R/\sqrt{\alpha'}$, or $\nu <1-\lambda ^4/4$.
which comes
from the requirement that the expression under the square root is
positive.
 
The dependence $E(U_0)$ derived from
(\ref{action}) is
\be
\label{energy}
E(U_0)=
\frac{U_0}{2\pi}\int_{1}^{\nu/\lambda}
\frac{y^2\sqrt{1-\lambda ^4 y^4}dy}
{ \sqrt{(y^4-1)(1-\lambda ^4(y^4+1))}}.
\ee

For small $\lambda$ from (\ref{L1}) we get
\be
\label{LU}
\frac{LU_0}{R^2}=2A+\lambda^4 C,
\ee
where
\be
A=\int_1^{\infty}{\frac{dy}{y^2\sqrt{y^4-1}}}=
\frac{\sqrt{2}\pi^{3/2}}{\Gamma(1/4)^2},
\ee
and $C$ is a constant, which could be determined by analyzing
the asymptotic behavior of the elliptic integrals.

Performing the renormalization in the expression for the energy 
as in \cite{Mald2}
and noting that the expression 
(\ref{energy}) includes the cutoff
we get
\be
\label{fe}
E(U_0)=U_0(c_0+\lambda^4 c_1)
\ee
where $c_0$ and $c_1$ are constants,
determined from (\ref{energy}). Therefore we get
\be
E(L)=2Ac_0 \cdot \frac{R^2}{L}+
(32 c_1 A^5+16A^4 c_0) \cdot \frac{\alpha'^2 R^6}{L^5}.
\label{EL}
\ee

\bigskip

Let us make a remark conserning 
the behavior of the string in the exact
3-brane solution (\ref{D3}).
The action (\ref{sa}) takes the form:
\be
S=\frac{T}{2 \pi} \int dx \sqrt{(\partial_x U)^2 +V(U)}
\label{FA}
\ee
where
\be
\label{pot}
V(U)=\frac{U^4}{R^4+\alpha ^{'2} U^4}.
\ee
The distance quark/anti-quark 
and the energy of the configuration are
defined by relations
\be
\label{LF}
\frac{L}{2}=
 \sqrt{V(U_0)}
\int_{U_0}^{U_{max}}
\frac{dU}{\sqrt{V(U)\cdot (V(U)-V(U_0))}},
\ee
\be
\label{FE}
E=
\frac{1}{2\pi}\int_{U_0}^{U_{max}}
\frac{\sqrt{V(U)}dU}{ \sqrt{V(U)-V(U_0)}}.
\ee

For potential (\ref{pot}) one has
\be
\label{fL}
\frac{LU_0}{R^2}=2A+\frac{2\alpha ^{' 2}U_0^4}{R^4}B,
\ee
\be
\label{fE}
E=\frac{U_0}{2\pi}\sqrt{1+\frac{\alpha ^{' 2}U_0^4}{R^4}}B,
\ee
where
\be
\label{intA}
A=\int_{1}^{U_{max}/U_{0}}{\frac{dy}{y^2\sqrt{y^4-1}}},
\ee
\be
\label{intB}
B=\int_{1}^{U_{max}/U_{0}}{\frac{y^2dy}{\sqrt{y^4-1}}},
\ee

We assume that the constant $B$ is renormalized as in \cite{Mald2}. From 
(\ref{fL}) and (\ref{fE}) one gets an expression for the energy
for large $L$:
\be
\label{fres}
E=\frac{R^2AB}{\pi L}[1+\frac{\alpha '^2 R^4}{L^4}8A^3(A+2B)]
\ee

If one assumes that for integrals (\ref{intA}) and (\ref{intB})
an N-dependence
cutoff is made so that
\be
\label{spec}
A=\frac{a}{R^2} , ~~~~~B=bR^2,
\ee
where $a$ and $b$ are N-independent, then one gets
\be
\label{E-sp}
E=\frac{abR^2}{\pi L}[1+\frac{\alpha '^2 }{L^4}(\frac{8a^4}{R^4}
+8a^3b)]
\ee
The first and the third terms can be interpreted as an interaction of
quarks
and the second term as an interaction between mesons.

\bigskip

I would like to thank the ICTP in Trieste and the organizers 
of the  Spring School on Non Perturbative Aspects of String Theory and
Supersymmetric Gauge Theories for the kind hospitality while this work
was done and for financial support.

\end{document}